\documentclass{article}
\usepackage{epsfig}
\usepackage{aip}

\newtheorem{theorem}{Theorem}
\newtheorem{acknowledgement}[theorem]{Acknowledgement}

\input{tcilatex}

\begin{document}

\title{\textbf{Matter fields from a decaying background }$\Lambda $\textbf{\ vacuum.%
}}
\author{Manasse R. Mbonye \\
\textit{Michigan Center for Theoretical Physics}\\
\textit{Physics Department, University of Michigan, Ann Arbor, Michigan 48109%
}}
\date{}
\maketitle

\begin{abstract}
We suggest an alternative framework for interpreting the current state of
the visible universe. Our approach is based on a dynamical ``Cosmological
Constant'' and the starting point is that a decaying vacuum produces matter.
By assuming inflation and big bang nucleosynthesis we can solve for the
present fractional densities of matter $\Omega _{m,0}$ and vacuum $\Omega
_{\Lambda ,0}$ in terms of only one parameter which we call the vacuum
domination crossing redshift, $z_{c}$. We put constraints on $z_{c}$ to
obtain a universe that is presently vacuum dominated and with characteristic
densities consistent with observations. The model points to the possible
existence of newly formed dark matter in the inter-cluster voids. We argue
that some of this matter could be accreting onto clusters through the
latter's long range gravitational potentials. If so, then cluster dark
matter halos may not manifest clear cut-offs in their radial density
profiles. Furthermore, if a substantial amount of this newly produced matter
has already drained onto the clusters, then the CMB power spectrum may favor
lower dark matter density values than is currently observed bound in the
clusters. A final feature of our approach relates to the combined effect of
the matter production by a decaying vacuum and the different rates at which
matter and the vacuum will dilute with the scale factor. Such combination
may create conditions for a universe in which the vacuum and matter
densities dilute and evolve towards comparable amplitudes. In this sense the
model offers a natural and conceptually simple explanation to the
Coincidence Problem.
\end{abstract}

\section{Introduction}

One of the most capricious and worrisome parameters in the entire history of
physics is the so-called cosmological constant, $\Lambda $. Ever since 1917
when Einstein first introduced it in Physics (through general relativity
(GR)) the cosmological constant has come and gone, only to come back again
and again. To date, observation and theory still disagree on the value of
this parameter by a staggering 120 orders of magnitude. It is the largest
discrepancy ever recorded in physics!

It is natural to associate the cosmological constant with a background
vacuum energy density, $\rho _{\Lambda }=\frac{\Lambda }{8\pi G}$. Recent
observations$^{1}$ suggest the present vacuum energy density, $\rho
_{\Lambda \text{ }}^{\left( 0\right) }$, to be of the same order as the
present matter density, $\rho _{m}^{\left( 0\right) }$, of the universe,
i.e. $\rho _{\Lambda \text{ }}^{\left( 0\right) }$ $\sim 10^{-30}\
g\,cm^{-3} $. On the other hand, in particle physics, vacuum energy results
from various sources, with each source contribution being much larger than
the upper limits of the total value given by observations. Quantum Field
Theory (QFT) calculations estimate$^{2}$ the background vacuum energy as $%
\langle \rho _{vac}\rangle \ \sim \;10^{94}\,g\,cm^{-3}$. It is this
discrepancy of more than 120 orders of magnitude between the observed and
the calculated values that constitutes the current Cosmological Constant
Problem. The problem can be stated in two parts: first, (i) why is the
observed $\Lambda $ so small and (ii) not zero? Secondly, why is the
associated energy density so \textit{coincidentally} close to the matter
energy density presently if the two were originally so disparately different
as predicted by particle physics? One possible explanation may be that
matter and the vacuum may be engaged in some kind of non-trivial
``communication''.

A number of attempts to explain the Cosmological Constant Problem have been
suggested since the 1980s. Most of these attempts are based on the
physicist's need to hold on to both results from theory and observation,
albeit (in this case) at different epochs of the universe. In general, this
view logically necessitates that the cosmological constant take on dynamical
attributes. There are two different traditional approaches usually taken in
addressing the problem. Particle physicists, for example, seek for
mechanisms to cancel or at least suppress $\Lambda $ (as in Coleman's
wormhole approach$^{3}$ or by inventing compensating fields in the
Langrangian). To date, however, there is no known symmetry argument that
justifies a vanishing (or worse still, an \textit{almost} vanishing) $%
\Lambda $.

The second approach makes use of concepts from general relativity. It is
also the approach on which the rest of this discussion is based. The
cosmological constant is known to have units of inverse length squared.
However, general relativity offers no associated length scale. One possible
natural scale, though, relates to the cosmological scale factor $a\left(
t\right) $ that appears in the Friedmann models. As it is, $a\left( t\right) 
$ is cosmologically time dependent and so, traditionally, not favored by GR
as the correct length scale for $\Lambda $. In Section 3 we will point out
that a time-dependant $\Lambda $ is not incompatible with the requirements
of GR provided we adopt a more generalized energy conservation approach.
With this one finds that it is precisely such time dependence that may offer
an exit window for the dynamical evolution of $\Lambda $ from its
QFT-calculated value to its presently observed one. The answer to the
Cosmological Constant Problem may be that simple. There is extensive
literature written on the Cosmological Constant Problem and for a thorough
review$^{2}$ see e.g. the work of Weinberg,1989 and that of Carrol, 2000.

In the 1980s several models \ sprang up$^{4}$ relating $\Lambda $ to the
scale factor $a\left( t\right) $ in the form $\Lambda \propto a\left(
t\right) ^{-2}$. It was not until 1990, however, that Chen and Wu$^{5}$
justified the functional form of such models based on general arguments from
quantum cosmology. Chen and Wu observed that one could always write $\Lambda 
$ in terms of $\left( m_{pl}\right) ^{4}$, ($m_{pl}$ is the Planck mass),
and some dimensionless product of quantities. Thus $\Lambda \propto \frac{1}{%
l_{pl}^{2}}\left( \frac{l_{pl}}{a\left( t\right) }\right) ^{n}$. Here $%
l_{pl} $ is the Planck length and $n$ is some number and $\hbar =c=1$. To be
consistent with GR which, as a classical theory, contains no $\hbar $, it is
necessary to set $n=2$. Later, Calvalho et al$^{6}$ suggested a dependence
of the form $\Lambda \propto \frac{1}{l_{pl}^{2}}\left( \frac{t_{pl}}{H}%
\right) ^{-n}$, where $t_{pl}$ and $H$ are the Planck time and the Hubble
(time) parameter, respectively. A more general model$^{7}$ combines the
above two to take the form $\Lambda \left( t\right) =\lambda _{1}a\left(
t\right) ^{-2}+\lambda _{2}H^{2}+\lambda _{0}$, where $\lambda _{0}$, $%
\lambda _{1}$ and $\lambda _{2}$, are adjustable parameters. In our
discussion we shall make use of the model originally suggested by \"{O}zer
and Taha$^{8}$ in which 
\begin{equation}
\Lambda \left( t\right) =\lambda _{1}a\left( t\right) ^{-2}+\lambda _{0}. 
\tag{1.1}
\end{equation}
In such a model the energy density is always decreasing as the system
continuously seeks a stable configuration. We shall therefore often refer to
this system as a dynamical false vacuum.

In astrophysics and cosmology, two of the traditional consequences of a
cosmological constant are the influence on the global geometry of the
universe and the redshift dependence of the luminosity-distance relation for
standard candles. It is, however, natural to ask whether there may be other
interesting consequences. In this paper we propose (as an extension to the
standard model of cosmology) an approach based on a dynamical cosmological
constant $\Lambda \left( t\right) $. We use the approach to derive several
of the cosmological quantities, including the present fractional densities
of matter $\Omega _{m,0}$ and vacuum $\Omega _{\Lambda ,0}$ contents in the
universe, in terms of one parameter we call the vacuum domination crossing
redshift (VDCR), $z_{c}$. On imposing constraints on $z_{c}$ we obtain a
universe that is presently vacuum dominated and with characteristic
densities consistent with observations. We discuss the important features of
this framework.

The remainder of this paper is arranged as follows. In Section 2 we
introduce a dynamical $\Lambda $\ and discuss its features in a cosmological
setting. We write down the general relativistic equations for a decaying
cosmological vacuum, discuss the conditions they satisfy, and the
implications for energy conservation in GR. In Section 3 these equations are
solved subject to constraints we introduce on the equation of state for an
interacting and dynamical vacuum. We discuss the asymptotic features of such
an equation of state and what they reveal about the possible dynamical
evolution of the universe. In Section 4 we focus our attention on the
implication of our results. We obtain cosmological parameters as functions
of one parameter and establish their consistency with observations. We
suggest tests of our model. Section 5 concludes the paper.

\section{\ A Dynamical vacuum}

\subsection{The cosmological setting}

The dynamics of the universe is described by the Einstein theory. Starting
from the $tt$ component of the Einstein equations (2.7) one obtains the
Friedmann equation which asserts that the (expansion $H$ of the) universe is
driven by the fields therein. Thus 
\begin{equation}
H^{2}=\left( \frac{\dot{a}\left( t\right) }{a\left( t\right) }\right) ^{2}=%
\frac{8\pi G}{3}\rho _{m}+\frac{\Lambda }{3}-\frac{k}{a\left( t\right) ^{2}},
\tag{2.1}
\end{equation}
where $\rho _{m}$ is the density of the matter fields, $k=\left\{
-1,0,1\right\} $, is the curvature constant. Further, the acceleration of
the universe is described by the Raychaudhuri equation, 
\begin{equation}
\frac{\ddot{a}\left( t\right) }{a\left( t\right) }=-\frac{4\pi G}{3}\left(
\rho +3p\right) ,  \tag{2.2}
\end{equation}
where $p=p_{m}+p_{\Lambda }$ is the pressure of all the fields in the
universe. Here and henceforth the subscripts $m$ and $\Lambda $ refer to the
matter and vacuum energy quantities, respectively.

The energy density associated with the cosmological constant in (1.1) can be
written as $\rho _{\Lambda }=\frac{\Lambda }{8\pi G}=\frac{\lambda
_{1}a\left( t\right) ^{-2}+\lambda _{0}}{8\pi G}$. Following \"{O}zer$^{9}$
we shall write $\rho _{\Lambda }$ in terms of the present values of the
scale factor $a\left( t_{0}\right) $ and matter component of the energy
density $\rho _{m}^{\left( 0\right) }$. (In our notation, the
subscript/superscript 0 will refer to the present value of a quantity). Then
using (1.1) we can write $\rho _{\Lambda }$ as 
\begin{equation}
\rho _{\Lambda }=\left[ \beta \left( \frac{a\left( t_{0}\right) }{a\left(
t\right) }\right) ^{2}+\alpha \right] \rho _{m}^{\left( 0\right) }, 
\tag{2.3}
\end{equation}
where we set $\alpha \rho _{m}^{\left( 0\right) }=\frac{\lambda _{0}}{8\pi G}
$ and $\beta a_{0}^{2}\rho _{m}^{\left( 0\right) }=\frac{\lambda _{1}}{8\pi G%
}$. Here $\alpha ,\beta \geq 0$ to ensure a positive energy density $\rho
_{\Lambda }$ from $\Lambda $. With a substitution of (2.3) into (2.1), the
Friedmann equation becomes \ 
\begin{equation}
H^{2}=\frac{8\pi G}{3}\rho _{m}\left( t\right) +\frac{8\pi G}{3}\alpha \rho
_{m}^{\left( 0\right) }-\frac{\left( k-\frac{8\pi G}{3}\beta a_{0}^{2}\rho
_{m}^{\left( 0\right) }\right) }{a\left( t\right) ^{2}}  \tag{2.4}
\end{equation}
As one notes from (2.4) such a dynamical $\Lambda $ will introduce, in the
theory, an effective curvature $k_{eff}$ given by

\begin{equation}
k_{eff}=k-\frac{8\pi G}{3}\beta \rho _{m}^{0}a\left( t_{0}\right) ^{2}. 
\tag{2.5}
\end{equation}
This result shows that with a dynamical $\Lambda $ the evolution of a
universe could depend on $k_{eff}$ instead of $k$ and hence on the (manner
of ) decay of $\Lambda $. For example with $\left( k_{eff}-k\right) \neq 0$,
a flat $\left( k=0\right) $ universe could effectively be dynamically open.
Further, a closed $\left( k=1\right) $ universe could be effectively flat or
even effectively open.

\subsection{General Relativistic Field Equations}

In this approach we consider the dynamical evolution of a self-gravitating
perfect cosmic fluid in a background vacuum with a variable and positive
cosmological constant, $\Lambda $. The total energy momentum tensor $T_{\mu
\nu }$ for all the fields is given by 
\begin{equation}
T_{\mu \nu }=\tilde{T}_{\mu \nu }+\rho _{\Lambda }g_{\mu \nu }=\left[ \rho +p%
\right] v_{\mu }v_{\nu }-pg_{\mu \nu },  \tag{2.6}
\end{equation}
where $\tilde{T}_{\mu \nu }$ is the matter part of $\ T_{\mu \nu }$, $\rho
=\rho _{m}+\rho _{\Lambda }$,\ $p=p_{m}+p_{\Lambda }$ and $v_{\mu }$ is the
4-velocity. In contrast to most other treatments in the literature we shall
find it useful to retain all the pressure terms $p_{m}$ and $p_{\Lambda }$
in the problem, and specialize to particular cases later. We suppose that $%
T_{\mu \nu }$ will satisfy Einstein field equations 
\begin{equation}
G_{\mu \nu }=8\pi G\tilde{T}_{\mu \nu }+\Lambda g_{\mu \nu },  \tag{2.7}
\end{equation}
where $G_{\mu \nu }$ is the Einstein tensor. The Bianchi identities $%
G_{\;\;;\mu }^{\mu \nu }=0$ imply that $T_{\mu \nu }$ is solenoidal, $\left(
T_{\;\;;\mu }^{\mu \nu }=0\right) $ and yield the conservation law $v_{\mu
}T_{\;\;;\nu }^{\mu \nu }=0$ which in our case in (2.7) takes the form $%
v^{\alpha }\rho _{;\alpha }+\left( \rho +p\right) v_{\;;\alpha }^{\alpha }=0$%
. More explicitly we can write this as 
\begin{equation}
\left[ \dot{\rho}_{m}+\left( \rho _{m}+p_{m}\right) \theta \right] +\left[ 
\dot{\rho}_{\Lambda }+\left( \rho _{\Lambda }+p_{\Lambda }\right) \theta %
\right] =0,  \tag{2.8}
\end{equation}
where $\theta =v_{\;;\alpha }^{\alpha }$ is the expansion parameter and $%
\dot{\rho}=v^{\alpha }\nabla _{\alpha }\rho $.

\subsection{Energy conservation in GR}

In the above scenario, we suppose that while the total energy is globally
conserved, matter and the vacuum can communicate. In particular, we suppose
that the vacuum can decay into matter\textit{. }\ This state of affairs
deserves commenting on.

In Einstein's theory, expressed in equation (2.7), the Einstein tensor $%
G_{\mu \nu }$ is conserved ($G_{\;\nu ;\mu }^{\mu }=0$)\ through the Bianchi
identities. Further, energy conservation requirements usually imply for the
matter fields that $\tilde{T}_{\,\,\nu ;\mu }^{\mu }=0$. On the other hand,
one can consider the cosmological term $\Lambda g_{\mu \nu }$ as part of the
energy momentum tensor $T_{\mu \nu }$. Then what we only need to be
conserved is the total $T_{\mu \nu }$. In this work, we take this latter
approach and relax the usual conservation conditions applied separately on
the matter fields and vacuum energy, respectively. We only demand a more
general or collective conservation law, $T_{\;\,\,\,\nu ;\mu }^{\mu }=0$,
(where $T_{\mu \nu }=\tilde{T}_{\mu \nu }+\frac{\Lambda }{8\pi G}g_{\mu \nu
} $). As a result, while the total energy is globally conserved, one can
still have source terms in individual contributions. This implies that one
can, in general, introduce a tensor $\Lambda _{\mu \nu }=\frac{\Lambda
\left( r,t\right) }{8\pi G}g_{\mu \nu }$ such that 
\begin{equation}
\tilde{T}_{\,\;\nu ;\mu }^{\mu }=a_{\nu }=-\Lambda _{\;\,\nu ;\mu }^{\mu } 
\tag{2.9}
\end{equation}
in which, clearly, the source of matter creation $a_{\nu }$ is the vacuum.
This generalized energy conservation which suggests a deeper
inter-relationship between spacetime and energy is also consistent with the
Bianchi identities' requirement that $G_{\;\nu ;\mu }^{\mu }=0$. General
relativity puts no further conditions on the form of the total energy
momentum tensor $T_{\mu \nu }$ other than that it must be conserved.

Such conceptions are not new and go back to Rastall's generalization of the
Einstein theory$^{10}$. Rastall's only constraint was that the vectors $%
a_{\nu }$ had to vanish in flat spacetime. This led him to postulate that
the $a_{\nu }$ should depend completely on the spacetime curvature. Such
dependence, in turn, introduces a spacetime curvature dependence in the
individual $T_{\mu \nu }$ components. As we noted in (2.5), the $a\left(
t\right) ^{-2}$ dependence of $\Lambda $ does indeed induce spatial
curvature on spacetime and in doing so modifies the Friedmann cosmology.
Conversely, recovering a dynamical $\Lambda $ assumes$^{11}$
non-conservation of the interacting individual fields (2.9). This suggests
then that it is possible to utilize a classical theory of gravity such as
general relativity to discuss the macroscopic aspects of matter production
by a vacuum. In this respect such a classical approach is to its would-be
quantum counterpart what analogously thermodynamics is to statistical
mechanics.

An important consequence of such an approach is that one can discuss the
thermodynamics of the situation in the general context of general
relativity. A multi-component energy momentum tensor such as $\tilde{T}_{\mu
\nu }+\frac{\Lambda }{8\pi G}g_{\mu \nu }$ for which the components are
allowed to interact constitutes a system akin to a closed multi-component
thermodynamic system and one whose adiabaticity is (in this case) guaranteed
through the Bianchi identities by $G_{\;\nu ;\mu }^{\mu }=0$. One notes$%
^{12} $ that the $a_{\nu }$ do, in fact, relate to the matter creation rate
and are also the source of entropy. As a result one can build up the
thermodynamics of this system by describing the particle creation current $%
N^{\alpha }$ and the entropy current $S^{\alpha }$ in terms of the source
term in $a_{\nu }$ (2.9). One then has that $N_{\;;\alpha }^{\alpha
}=Cv^{\alpha }a_{\alpha }=\psi $, where $C$ is some constant and $\psi >0$
is the particle source. This can be written more transparently from $%
N=nv^{\alpha }$ (where $n$ is the particle number density) as 
\begin{equation}
\dot{n}+n\theta =\psi ,  \tag{2.10}
\end{equation}
and the entropy current is given by 
\begin{equation}
S^{\alpha }=sv^{\alpha }=\sigma N^{\alpha }=n\sigma v^{\alpha },  \tag{2.11}
\end{equation}
where $s$ is the entropy density and $\sigma $ is the entropy per particle,
i.e. the specific entropy.

\section{Vacuum decay and matter fields}

\subsection{Solution for matter fields}

\smallskip For the above reasons we assume global conservation of energy in
the universe while allowing sources and sinks in its components. In the FRW
models, the source of the four-velocity is the Hubble parameter $%
v_{\;;\alpha }^{\alpha }=$ $\theta =3\frac{\dot{a}}{a}=3H$ and the energy
conservation equation (2.8) gives rise to the first law of thermodynamics 
\begin{equation}
d\left[ \rho a\left( t\right) ^{3}\right] +pd\left[ a\left( t\right) ^{3}%
\right] =0.  \tag{3.1}
\end{equation}
We shall suppose that the vacuum satisfies a barotropic equation of state of
the form 
\begin{equation}
p_{\Lambda }=w\rho _{\Lambda },  \tag{3.2}
\end{equation}
where $-1\leq w<0$. Specifically, we adapt the view that $\Lambda ^{\mu \nu
} $does, itself, constitute a bi-component fluid. The constant $\left(
\alpha \right) $\ part (see 2.3) has the usual vacuum equation of state with 
$w=-1$\ while the dynamical part has $w=w_{\beta }$, so that $w=\left\{
\left( w_{\alpha }=-1\right) ,w_{\beta }\right\} $\ with $w_{\beta }$\ to be
determined later.\textit{\ }Thus the dynamical part of $\Lambda $\ is a
fluid with an energy momentum tensor of the form $diag(\rho _{\beta
},w_{\beta }$\thinspace $\rho _{\beta },\,w_{\beta }\rho _{\beta
},\,w_{\beta }\rho _{\beta })$. We will argue later that $w_{\beta }$\ has a
dependence on the matter fields (among other things) such that in the limit\ 
$\rho _{m}\longrightarrow 0$\ then $w_{\beta }\longrightarrow -1$. In this
limit this tensor is proportional to the Minkowski metric $diag(-1,1,1,1)$\
with the result that in a locally inertial frame it is Lorentz invariant.

Further, we assume that the regular matter fields satisfy the usual $\gamma $%
-law equation of state. 
\begin{equation}
p_{m}=\left( \gamma -1\right) \rho _{m},  \tag{3.3}
\end{equation}
with $\gamma =\left\{ 1,\frac{4}{3}\right\} $. As we soon find, the
incremental matter produced from the $\Lambda -$vacuum decay does not
generally satisfy this law.

With the above equations of state (3.2) and (3.3) along with use of (2.3),
(3.1) becomes

\begin{equation}
\begin{array}{c}
d\left\{ \left[ \rho _{m}\left( t\right) \right] a\left( t\right)
^{3}\right\} +\left( \gamma -1\right) \rho _{m}\left( t\right) d\left[
a\left( t\right) ^{3}\right] \\ 
+ \\ 
d\left[ \left( \beta \frac{a\left( t_{0}\right) ^{2}}{a\left( t\right) ^{2}}%
+\alpha \right) \rho _{m}^{\left( 0\right) }a\left( t\right) ^{3}\right]
+\left( w_{\beta }\beta \frac{a\left( t_{0}\right) ^{2}}{a\left( t\right)
^{2}}+w_{\alpha }\alpha \right) \rho _{m}^{\left( 0\right) }d\left[ a\left(
t\right) ^{3}\right] =0.
\end{array}
\tag{3.4}
\end{equation}
Equation (3.4) can be integrated to the present $\left( t=t_{0}\right) $ to
yield the solution 
\begin{equation}
\begin{array}{c}
\rho _{m}\left( t_{0}\right) a\left( t_{0}\right) ^{3\gamma }-\rho
_{m}\left( t\right) a\left( t\right) ^{3\gamma }= \\ 
\\ 
\left( \frac{\beta }{3\gamma -2}\right) \left( 1+3w_{\beta }\right) \left[ 
\frac{a\left( t_{0}\right) ^{2}}{a\left( t\right) ^{2}}a\left( t\right)
^{3\gamma }-a\left( t_{0}\right) ^{3\gamma }\right] \rho _{m}^{\left(
0\right) }+\frac{\alpha }{\gamma }\left( 1+w_{\alpha }\right) \left[ a\left(
t\right) ^{3\gamma }-a\left( t_{0}\right) ^{3\gamma }\right] \rho
_{m}^{\left( 0\right) }
\end{array}
\tag{3.5}
\end{equation}
This solution gives the matter produced in the time interval $\Delta
t=t_{0}-t$ during which the scale factor grows from $a\left( t\right) $ to $%
a\left( t_{0}\right) $. The density $\rho _{m}\left( t\right) $ then evolves
as 
\begin{equation}
\begin{array}{c}
\rho _{m}\left( t\right) =-\beta \left( \frac{1+3w_{\beta }}{3\gamma -2}%
\right) \left[ \frac{a\left( t_{0}\right) ^{2}}{a\left( t\right) ^{2}}-\frac{%
a\left( t_{0}\right) ^{3\gamma }}{a\left( t\right) ^{3\gamma }}\right] \rho
_{m}^{\left( 0\right) }-\frac{\alpha }{\gamma }\left( 1+w_{\alpha }\right) 
\left[ 1-\frac{a\left( t_{0}\right) ^{3\gamma }}{a\left( t\right) ^{3\gamma }%
}\right] \rho _{m}^{\left( 0\right) }+\frac{a\left( t_{0}\right) ^{3\gamma }%
}{a\left( t\right) ^{3\gamma }}\rho _{m}^{\left( 0\right) }.
\end{array}
\tag{3.6}
\end{equation}
This simple result and its implications will be at the core of the remaining
discussion. For a matter conserving universe, the right hand side of \ (3.5)
should vanish. As we soon find this is not generally the case, with the
exception of the limiting and unlikely (see next subsection) scenario when $%
w_{\beta }=-\frac{1}{3}$. By setting $\gamma =1$ or $\gamma =\frac{4}{3}$,
respectively, one can easily verify that equation (3.6) yields matter fields
in the matter dominated era or radiation fields in the radiation era.

\subsection{\protect\smallskip Field dilution}

In the remaining part of this section, we develop a framework for discussing
the dynamics of the fields we have considered so far. The treatment is not
rigorous and only the asymptotic behavior of such dynamics is discussed.
Nevertheless, the results from this treatment will be useful in as far as
they provide us with an insight in the possible dynamics of the universe.
More particularly we shall use such results in the next section to relate
our model to observations.

As the universe expands the densities of the matter fields dilute as $%
a(t)^{-3\gamma }=a(t)^{-3(1+\breve{w})}$, where $\breve{w}=\left\{ 0,\frac{1%
}{3}\right\} $. On the other hand, as we have shown above a decaying
cosmological constant produces matter/entropy, with a particle current $%
N^{\alpha }$ given by $N_{\;;\alpha }^{\alpha }=-Cv^{\alpha }\Lambda
_{\;\;\alpha ;\beta }^{\beta }=\psi $. As a source of matter fields $\Lambda 
$ will suffer a further energy density deficit since $\Lambda _{\;\,\nu ;\mu
}^{\mu }=a_{\nu }\neq 0$. This energy loss by the vacuum creates some extra
(dimension-like) degree(s) of freedom in the dilution law thus enhancing
further dilution of the vacuum energy density. It follows then that such a
field dilutes as $a(t)^{-\left( 3+\varepsilon \right) (1+w)}$, where the
dimensionless parameter $\varepsilon \geq 0$ (still to be defined) accounts
for the extra degree(s) of freedom.

We stated before that the vacuum energy density, $\rho _{\Lambda }=\left[
\beta \left( \frac{a\left( t_{0}\right) }{a\left( t\right) }\right)
^{2}+\alpha \right] \rho _{m}^{\left( 0\right) }$, is a multi-component
field. When $w=w_{\alpha }=-1$ so that one recovers the usual vacuum
equation of state $p_{\alpha }=-\rho _{\alpha }$ one can verify from the $%
a(t)^{-\left( 3+\varepsilon \right) (1+w)}$ dilution law that $\Lambda
_{\alpha }$ is, indeed, a constant. The second terms in equations (3.6) will
then vanish, showing as expected, that the constant part of $\Lambda $ plays
no role in matter creation. This reduces equation (3.6) to 
\begin{equation}
\rho _{m}\left( t\right) =-\left( \frac{\beta }{3\gamma -2}\right) \left(
1+3w_{\beta }\right) \left( \frac{a\left( t_{0}\right) ^{2}}{a\left(
t\right) ^{2}}-\frac{a\left( t_{0}\right) ^{3\gamma }}{a\left( t\right)
^{3\gamma }}\right) \rho _{m}^{\left( 0\right) }+\frac{a\left( t_{0}\right)
^{3\gamma }}{a\left( t\right) ^{3\gamma }}\rho _{m}^{\left( 0\right) } 
\tag{3.7}
\end{equation}

On the other hand, to recover the $\Lambda \sim \frac{1}{a(t)^{2}}$ scaling
from the dilution law clearly requires that we set $w=w_{\beta }=-\frac{%
1+\varepsilon }{3+\varepsilon }$. This condition imposes on the dynamical
part of $\Lambda $ an equation of state of the form 
\begin{equation}
p_{\beta }=-\left( \frac{1+\varepsilon }{3+\varepsilon }\right) \rho _{\beta
}.  \tag{3.8}
\end{equation}

From equation (3.8) we note that for large values of $\varepsilon $
(assuming such values exist) $p_{\beta }$ is a slowly varying function of $%
\varepsilon $. This suggests that the presence of matter fields may have an
influence in the dynamical evolution of vacuum energy density. To see this
we assume that $\varepsilon $ is a function of a dimensionless quantity $%
\theta $ where $\theta \left( t\right) =\frac{\rho _{\beta }\left( t\right) 
}{\rho _{m}\left( t\right) }$ is the vacuum to matter ratio at any time $t$.
We suppose that the matter creation window $\varepsilon \left( \theta
\right) $ can be written as a polynomial in this ratio, $\varepsilon \left(
\theta \right) =\sum c_{n}\left[ \theta \left( t\right) \right] ^{n}$, where 
$n\geq 0$ and the $c_{n}$ are undetermined coefficients. For the purposes of
this discussion we need not know the exact functional form of $\varepsilon
=\varepsilon \left( \theta \right) $ as it is sufficient to only establish
the asymptotic limits. As the false vacuum energy density evolves to
dominate the universe so that $\rho _{\beta }>>\rho _{m}$ then the equation
of state (3.8) evolves so that $\stackunder{\varepsilon \rightarrow \infty }{%
\lim }p_{\beta }=-\rho _{\beta }$. It follows then that in the limit of
vanishing matter fields (and hence their gravitational effects) such an
interacting dynamical vacuum acquires the equation of state, $p_{\beta
}=-\rho _{\beta }$, of the traditional cosmological constant. This implies
that in a locally inertial frame the tensor, $\left( \Lambda _{\beta
}\right) _{\mu \nu }$ $=diag\left( \rho _{\beta },p_{\beta },p_{\beta
},p_{\beta }\right) $, associated with the dynamical part of the vacuum,
will be proportional to the Minkowski metric $diag\left( -1,1,1,1\right) $.
As a result, equation (3.8) constitutes the pressure component of an
asymptotically locally Lorentz invariant tensor field. Thus a dynamical $%
\Lambda $ is not incompatible with the general requirements of general
relativity.

Conversely, as the matter fields dominate so that $\theta \rightarrow 0$
then $\varepsilon \rightarrow 0$ and equation (3.8) shows that $\stackunder{%
\varepsilon \rightarrow 0}{\lim }p_{\beta }=-\frac{1}{3}\rho _{\beta }$.
Note that in this limit one recovers the usual equation of state for an
exotic fluid (such as that of randomly oriented cosmic strings). From
equation (3.4), however, it is clear that the limit $w_{\beta }=-\frac{1}{3}$
results in no matter creation. This\ has interesting implications. It
suggests that matter production by a decaying vacuum will, in general, be
suppressed in a matter dominated universe. Now, since our universe is known
to have been matter/radiation dominated until recently (see next section)
the inference is that any interesting incremental matter creation, beyond
big bang nucleosynthesis (BBN), could only have taken place after the vacuum
domination crossing redshift $z_{c}$ which (as we show soon) is long after
even the delayed form$^{13}$ of recombination.\textit{\ }This observation
will be useful later as we relate the results of our model with the
information imprinted on the cosmic microwave background (CMB) radiation
power spectrum.

As a window regulating both matter production and the vacuum dilution rate, $%
\varepsilon \left( \theta \right) $ may have an influence on the future
evolution of the universe. In our model, an increase in the vacuum to matter
ratio $\theta $ increases $\varepsilon $ which increases matter production.
A consequence of such incremental matter production is that the regular
matter dilution rate $a\left( t\right) ^{-3\gamma }$ will be apparently
offset towards the lower end. Moreover, conservation requirements imply (as
mentioned before) that the vacuum dilution be offset (towards the higher
end). This implies that there will be an average critical value of $%
\varepsilon \left( \theta \right) $ above which the amount of matter
produced by the decaying vacuum will, apparently, slow down the matter
dilution rate $a\left( t\right) ^{-3\gamma }$ enough that the now diluted
vacuum density starts to drop below the matter density. Recalling that $%
\varepsilon $ is a dimension-like degree of freedom this suggests a critical
average value of $\varepsilon \simeq \frac{1}{2}$ ($\varepsilon \simeq 1$
for radiation). Thus for $\varepsilon <\frac{1}{2}$ the rate of matter
production from the decaying vacuum would probably never be high enough to
significantly compromise the matter dilution. In that case $\theta $ would
then evolve towards some asymptotic value and the vacuum would operate from
there.

On the other hand, our arguments above suggest that for $\varepsilon >\frac{1%
}{2}$, the associated matter production rate and vacuum decay rate would
eventually result in a matter dominated universe. Clearly a high rate of
matter creation, in turn, suppresses $\varepsilon =\varepsilon \left( \beta
\right) $ sending the equation of state $p_{\beta }=-\left( \frac{%
1+\varepsilon }{3+\varepsilon }\right) \rho _{\beta }$ back towards the
matter dominated form, $\stackunder{\varepsilon \rightarrow 0}{\lim }%
p_{\beta }=-\frac{1}{3}\rho _{\beta }$. As we have seen, in the neighborhood
of this latter limit matter production is disfavored. Moreover, as matter
production becomes suppressed, the universe still continues to expand
redshifting the matter fields as $a\left( t\right) ^{-3}$ so that eventually
the vacuum begins to dominate again albeit at a lower energy density than
before. The result is a universe in which the vacuum and matter fields
densities would oscillate into each other and do so within a decaying
envelope. This may be the key to solving the Coincidence Problem. Moreover,
such a process would suggest a universe stable to vacuum induced runaway
accelerations. Further still, the enveloping damping behavior resulting from
vacuum dilution associated with matter creation would tend to evolve such a
universe towards some stable attractor state in the far future.

In this treatment we take the position that, at the present epoch of the
universe, $w_{\beta }$ is a slowly varying function of time. In the next
section as we seek to relate the results of our model with observation, we
adopt an average value of $\varepsilon =1$ for the decay window. This
results in a working equation of state 
\begin{equation}
p_{\beta }=-\frac{1}{2}\rho _{\beta }  \tag{3.9}
\end{equation}
for the dynamical part of the cosmological constant and which lies well
within the bounds allowed by (3.8). As we verify soon, this choice of
equation of state, will be justified by the consistency of its results with
observation.

In closing we mention that our result seems to bear a resemblance to
Quintessence$^{14}$. This result, however, differs from (Zlatev 1999) in
several respects. First the vacuum we consider is pointwise dynamical and
therefore persistently false. Secondly this decaying vacuum produces matter.
Moreover, the field involved is a \textit{perturbed} cosmological constant
whose associated energy momentum tensor becomes Lorentz invariant in the
(theoretical) absence of matter fields but whose equation of state may, in
general, be time dependent. Incidentally, such a locally inertial frame may
only exist in theory since once some matter already exists then the limit $%
\theta \longrightarrow 0$ may not exist in finite time.

\section{Implications}

In this section we shall point out the cosmological implications of the
preceding arguments and discuss how these implications can contribute to our
interpretation of the observed universe. The discussion is divided in three
parts. First we use these arguments to derive expressions for several
cosmological parameters. Secondly we seek to relate these latter results to
observations. Thirdly we point out some of the features our approach brings
to astrophysics and cosmology. We seek for a working testable model that
attempts to make sense of some of recent developments in observational
cosmology.

\subsection{Cosmological parameters}

Here and henceforth we write the fields in terms of the redshift $z=\frac{%
a_{0}}{a\left( t\right) }-1$ and set $\gamma =1$. With this and using
equation (3.9) the expression in equation (3.7) for the matter field density
becomes

\begin{equation}
\rho _{m}\left( z\right) =[1+z\left( 1-\frac{1}{2}\beta \right)
](1+z)^{2}\rho _{m}^{\left( 0\right) }.  \tag{4.1}
\end{equation}
In section 3 we established that the constant part of the vacuum energy
density in equation (2.3) does not take part in matter production and so for
the purposes of our discussion here we set $\alpha \approx 0$ and consider
the vacuum as fully dynamical, i.e. 
\begin{equation}
\rho _{\Lambda }=\beta \left( 1+z\right) ^{2}\rho _{m}^{\left( 0\right) }, 
\tag{4.2}
\end{equation}
and with a working equation of state given by (3.9). Thus as a fraction of
the critical density $\rho _{c}$ the present $\left( z=0\right) $ value of
the vacuum energy density becomes 
\begin{equation}
\Omega _{\Lambda ,0}=\frac{\rho _{\Lambda }}{\rho _{c}}=\beta \frac{\rho
_{m}^{\left( 0\right) }}{\rho _{c}}=\beta \Omega _{m,0},  \tag{4.3}
\end{equation}
where $\Omega _{m,0}$ is the current fractional contribution of the total
matter density.

Following the arguments developed in the previous section we assume matter
production sets in at the vacuum domination crossing redshift (VDCR), $%
z=z_{c}$. This is when the universe enters the vacuum-dominated phase. In
the later part of this section, we shall impose some constraints on $z_{c}$
as we establish that the universe is indeed currently vacuum dominated. The
VDCR is set by the condition $\rho _{m}\left( z\right) =\rho _{\Lambda
}\left( z\right) $. Using equations (4.1), and (4.2.) one finds\textbf{\ }%
that, in terms of $\beta $, 
\begin{equation}
z_{c}=\frac{2\left( \beta -1\right) }{2-\beta }.  \tag{4.4}
\end{equation}
Notice that the vacuum domination crossing redshift, $z_{c}$ puts
constraints on the free parameter $\beta ,$ which relates the vacuum to the
matter content in the universe. Clearly equation (4.4) constrains $\beta $
in the tight range $1<\beta <2$. Such constraints seem to be physically
justifiable. The lower bound on $\beta $, which translates to $z_{c}>0$, is
consistent with the observation$^{1}$ that vacuum energy already dominates
the universe. The upper bound value of $\beta $, on the other hand,
guarantees that there was matter (see equation (4.6)) before the present
vacuum domination so that only a fraction $\Delta \Omega _{m}$ of the total
matter has been produced since. A challenge for the model is to accommodate
the observation that before the vacuum domination $z>z_{c}$, there had to
have been enough pre-existing matter $\Omega _{m}\left( z>z_{c}\right) $\ to
produce structure$^{15}$. As we soon find, $\Omega _{m}\left( z>z_{c}\right) 
$\ should constitute most of the matter bound up in galaxy clusters.

The amount of matter produced since vacuum domination crossing contributes a
piece $\Delta \rho _{m}^{\left( 0\right) }$, to the present matter density $%
\rho _{m}^{\left( 0\right) }$, given by $\Delta \rho _{m}^{\left( 0\right)
}=\rho _{m}^{\left( 0\right) }-\rho _{m}\left( t\right) \left( 1+z\right)
^{-3}$. Using equation (4.1) to substitute for $\rho _{m}\left( z\right) $
one finds $\Delta \rho _{m}^{\left( 0\right) }=\frac{1}{2}\beta \rho
_{m}^{\left( 0\right) }\left( \frac{z_{c}}{1+z_{c}}\right) $. Along with
(4.4) one finally gets 
\begin{equation}
\Delta \Omega _{m}=\left( \frac{z_{c}}{2+z_{c}}\right) \Omega _{m,0}. 
\tag{4.5}
\end{equation}
Since equation (4.4) implies $z_{c}>0$, then equation (4.5) implies $\Delta
\Omega _{m}>0$, reaffirming that the decay of the vacuum does indeed result
in matter production. Consequently, the current matter content of the
universe is given by $\Omega _{m,0}=\Omega _{m}\left( z>z_{c}\right) +\Delta
\Omega _{m}$. The present partial density due to the matter that existed
before vacuum domination $\Omega _{m,old}$\ is given, with the use of (4.5),
by 
\begin{equation}
\Omega _{m,old}=\Omega _{m}\left( z>z_{c}\right) =\left( \frac{2}{2+z_{c}}%
\right) \Omega _{m,0}.  \tag{4.6}
\end{equation}

The issue of the form of matter the vacuum decays into is outside the scope
of this discussion and requires a quantum gravity approach. However, for a
vacuum currently with an associated energy density$^{1}$ as low as $\rho
_{\Lambda \text{ }}^{\left( 0\right) }$ $\sim 10^{-30}\ g\,cm^{-3}$, one can
easily rule out the possibility of regular baryonic matter as a decay
product. The contents of $\Delta \Omega _{m}$ can only be some kind of dark
matter. This dark matter should, on the other hand, be different from the
`old' (pre-vacuum domination) dark matter $\Omega _{dm,old}$ in $\Omega
_{m,old}$ in the sense that the conditions under which each is formed are
different. Since $\Delta \Omega _{m}$ is all dark matter it follows that $%
\Omega _{m,old}$ in (4.6) must contain all the baryonic matter $\Omega _{B}$
of the universe. This implies that the baryonic content of the universe
should solely be that predicted by big bang nucleosynthesis, BBN$^{16}$.
Moreover, $\Omega _{m,old}$ also contains the `old' dark matter $\Omega
_{dm,old}$ that was necessary to initiate gravitational instability (GI) in
the early universe$^{17}$. Thus $\Omega _{m,old}=\Omega _{B}+\Omega
_{dm,old} $.

And so with respect to the form of the created `new' dark matter $\Delta
\Omega _{m}$, the guess is on the standard candidates: neutrinos (possibly
cold/massive), axions and/or wimps and/or possibly some other (not yet
discovered) matter forms. The `new' matter could also take a form similar to
the Cold and Fuzzy Dark Matter recently suggested by Hu et al$^{18}$. This
latter form is somewhat appealing in that it suits the existing low energy
density conditions of the current decaying vacuum. In any case as implied in
equation (4.5) the matter in $\Delta \Omega _{m}$ has a positive energy
density. We also assume that, in the end, it is relatively cool and obeys
the standard energy conditions$^{19}$.

In the inter-galactic voids the matter production by the decay of $\Lambda
\left( t\right) $ is homogeneous, on a fixed spacetime hyperface. However,
in the neighborhood of galaxy clusters such matter products should tend to
respond to the former's long range gravitational potentials and, in time,
gravitate towards the galaxies and clusters, eventually serving to increase
the halo mass. Later we shall suggest a possible, observationally distinct,
consequence of this accretion process.

\subsection{The $D\Lambda -CDM$ model and observations}

In the remaining part of our discussion we introduce a model we call ``the
dynamical-$\Lambda $ and cold dark matter'' or $D\Lambda -CDM$, for short.
The model which is based on the preceding arguments of a dynamical $\Lambda $%
, also assumes inflation$^{20}$ and BBN$^{16}$. In the end, we seek for
consistency of the results of $D\Lambda -CDM$\ with observation by
constraining several parameters of cosmology, including the total matter
density $\Omega _{m,0}$ and the vacuum density $\Omega _{\Lambda }$.

\ \ \ The inflationary model of cosmology$^{20,21}$ predicts a spatially
flat universe with a critical density $\Omega _{tot}=\Omega _{\Lambda
,0}+\Omega _{m,0}=\Omega _{c}=1$. With this, one can now write the above
cosmological quantities in terms of only one parameter, namely the vacuum
domination crossing redshift, $z_{c}$. Thus the inflationary condition along
with (4.3) and (4.4) give the present matter density $\Omega _{m,0}$\ of the
universe as 
\begin{equation}
\Omega _{m,0}=\frac{2+z_{c}}{4+3z_{c}},  \tag{4.7}
\end{equation}
while the present value of the vacuum energy density $\Omega _{\Lambda ,0}$
is 
\begin{equation}
\Omega _{\Lambda ,0}=\frac{2\left( 1+z_{c}\right) }{4+3z_{c}}.  \tag{4.8}
\end{equation}
Further, for the present partial density contribution by the pre-vacuum
domination $\left( z>z_{c}\right) $ matter we have 
\begin{equation}
\Omega _{m,old}=\frac{2}{4+3z_{c}},  \tag{4.9}
\end{equation}
while that due the matter created from vacuum decay since is 
\begin{equation}
\Delta \Omega _{m}=\frac{z_{c}}{4+3z_{c}}.  \tag{4.10}
\end{equation}
Figure 1a shows how the relative magnitudes of the current vacuum density $%
\Omega _{\Lambda ,0}$, total matter density $\Omega _{m,0}$, the pre-vacuum
domination matter density $\Omega _{m,old}$ and the ``new'' matter density $%
\Delta \Omega _{m}$ depend on the redshift $z_{c}$\ at which the vacuum
begins to dominate the universe. Note that the curve for $\Omega _{m,0}$ is
simply the sum of the two curves $\Omega _{m,old}+\Delta \Omega _{m}$. Since
baryonic matter is ruled out in the present vacuum decay products, it is
convenient in our discussion to hold $\Omega _{B}$ fixed (Figure 1b) as we
study the changes in the total matter content resulting from the change in
the dark matter content. The present value of total dark matter $\Omega
_{dm,0}$ and its pre-vacuum domination $\left( z>z_{c}\right) $\ component $%
\Omega _{dm,old}$ can be written down, respectively, as $\Omega
_{m,0}-\Omega _{B}$ and $\Omega _{m,old}-\Omega _{B}$, where for the
baryonic component of the present energy density we take (Tytler 2000,
Burles 1998, Burles 1999) the big bang nucleothenthesis (BBN) central value
as $\Omega _{B}h^{2}=0.019\pm .0012$. Figure 1b shows the dependence on $%
z_{c}$ of\ $\Omega _{\Lambda ,0}$, $\Omega _{dm,0}$, $\Omega _{dm,old}$ and $%
\Delta \Omega _{m}$, assuming BBN and using$^{22}$ the prior $h=0.72\pm 0.08$%
. Note that in Figure 1b the gap between $\Omega _{\Lambda ,0}$ and $\Omega
_{dm,0}$ at $z_{c}=0$ is just $\left| \Omega _{B}\right| $. We seek to
constrain these parameters. \ The $D\Lambda -CDM$ model already predicts a
currently vacuum dominated universe, as shown by equations (4.7) and (4.8)
(see also Figure 1). However, the constraints we impose on $\Omega _{dm,old}$%
\ help to give us bounds on this and other quantities. In the forecoming
discussion our aim is not so much to focus on the precision of the data we
use than it is to demonstrate the usefulness of this framework in
interpreting such data. Consequently we shall often, in our discussion, just
make use of the central values of the results we quote.

In adiabatic inflationary models$^{17}$, structure formation is preceded by
gravitational instability (GI) in dark matter. Such instability results in
the formation of gravitational potentials into which baryonic matter is
eventually drawn. Under assumptions of Gaussian initial perturbations a
central value of the lower bound matter density of $0.2$ is usually quoted$%
^{23}$. We take this value of $\Omega _{m,old}=0.2$ and the associated $%
\Omega _{dm,old}\simeq 0.16(3)$ as our lower bound weak prior for the early
universe to generate structure. In our $D\Lambda -CDM$ model any such
initial condition will also affect the future of the dynamical universe in
as far as it pre-determines (see equation (4.8)) the vacuum domination
crossing redshift $\left( VDCR\right) $, $z_{c}$. Figure 1b shows that the
value $\Omega _{dm,old}=0.16$ is consistent with a vacuum domination redshit 
$z_{c}\simeq 2.0$. We take this redshift as our working upper bound value
for $z_{c}$. The shaded region (in Figure 1b) on the right of $z_{c}\simeq 2$
is therefore restricted (albeit weakly) by structure formation. As Figure 1b
shows, at this upper bound $z_{c}$ we obtain an upper bound vacuum density
value of $\Omega _{\Lambda ,0}\simeq 0.60$ and a corresponding lower bound
on total matter density of $\Omega _{m,0}\simeq 0.4$ (dark matter density of 
$\Omega _{dm,0}\simeq 0.36)$. We also obtain here an upper bound on the
matter created since this $VDCR$ to date as $\Delta \Omega _{m}\simeq 0.20$.
Now, we suppose that the created matter should slowly gravitate towards the
clusters. Based on measurements from cluster baryon fraction techniques$%
^{24} $ we assume a central value of matter density $\tilde{\Omega}%
_{m,0}\simeq 0.3\pm 0.1$ bound in clusters (i.e. dark matter density $\tilde{%
\Omega}_{dm,0}\simeq 0.26$) and estimate the upper bound for created matter
already accreted onto the clusters to be $\tilde{\Omega}_{dm,0}-\Omega
_{dm,old}\simeq 0.10$, leaving an estimate of the lower bound for
inter-cluster matter to be $\Omega _{dm,0}-\tilde{\Omega}_{dm,0}\simeq 0.10$.

\begin{figure}[htp]
\centering
\leavevmode
\includegraphics[scale=0.5,origin=c]{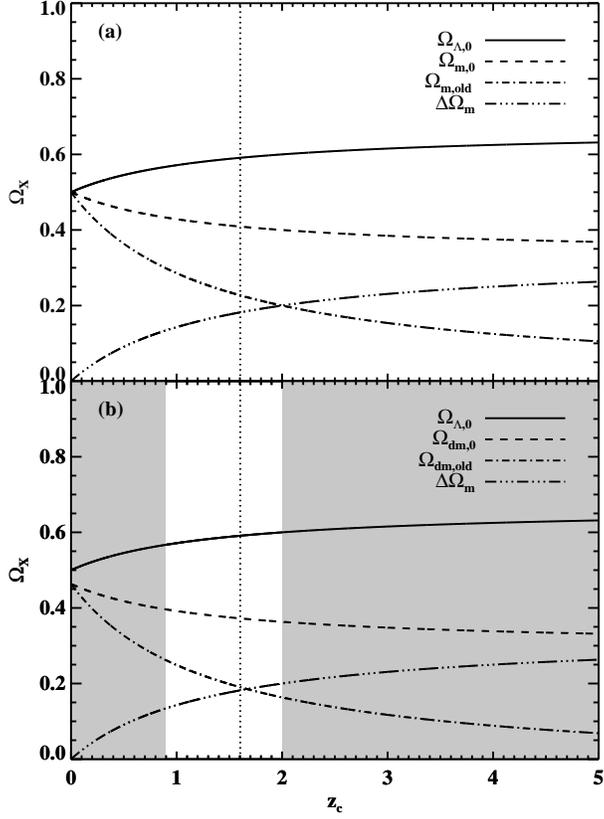}
\caption{ (a) Current relative magnitudes of vacuum density, $\Omega
_{\Lambda ,0}$, total density $\Omega _{m,0}$, the ``old'' matter density $%
\Omega _{m,old}$ and the ``new'' matter density $\Delta \Omega _{m}$ depend
on $VDCR$, $z_{c}.$ (b) The same dependence for dark matter $\Omega
_{dm,old} $ and $\Omega _{dm,0}$ with $\Omega _{B}$ fixed at the BBN value.
Using priors on $\Omega _{dm,old}$ to put bounds on $z_{c}$ we constrain all
the other quantities. The vertical dotted line gives results based on $%
\Omega _{dm,old} $ used to simulate recent (Wang 2001) best CMB fit. }
\end{figure}




At the other extreme it is possible that none of the decay products in $%
\Delta \Omega _{m}$ have yet deposited onto the bound matter in clusters. In
that case the value of dark matter bound in clusters$^{24}$, $\left( \tilde{%
\Omega}_{m,0}-\tilde{\Omega}_{B}\right) \simeq 0.26(3)$, would be solely
that due to the old dark matter $\Omega _{dm,old}\simeq 0.26(3)$. A look at
Figure 1b shows that\ such a value corresponds to a lower bound on $VDCR$ of 
$z_{c}\simeq 0.89$. This, in turn, yields a lower bound on the decay
products of $\Delta \Omega _{m}\simeq 0.13(3)$, an upper bound on total
matter $\Omega _{m,0}\simeq 0.43$ (dark matter $\Omega _{dm,0}\simeq 0.39(6)$%
) and a lower bound on vacuum energy of $\Omega _{\Lambda ,0}\simeq 0.57$.
In Figure 1b, the shaded region on the left is disfavored by these
constraints.

We can now summarize the constraints which, with the use of the selected
priors, our model imposes on the cosmological parameters.\textbf{\ }Recall
that central values of $\Omega _{dm,old}$ are used to choose the constraints
and as such the latter are only intended to be illustrative. Given these
assumptions, our model tells a story of a universe that started off with
dark matter that now contributes a partial density $0.16\leq \Omega
_{dm,old}\leq 0.26$. Recently, at a redshift of $0.89\leq z_{c}\leq 2.0$ the
universe became vacuum dominated. Since then the decaying vacuum has
produced matter. This new matter contribution presently amounts to $0.13\leq
\Delta \Omega _{m}\leq 0.20$ with as much as $0\leq \left( \tilde{\Omega}%
_{dm,0}-\Omega _{dm,old}\right) \leq 0.10$ already deposited on to the
clusters, leaving as much as $0.10\leq \left( \Omega _{dm,0}-\tilde{\Omega}%
_{dm,0}\right) \leq 0.13(3)$ to be inter-cluster unbound matter. The result
is a vacuum dominated universe with a vacuum density $0.57\leq \Omega
_{\Lambda ,0}\leq 0.60$\ and a total (including $\Omega _{B}$) matter
density $0.40\leq \Omega _{m,0}\leq 0.43$.

\subsection{$D\Lambda -CDM$ and CMB}

It is evident from this discussion of the $D\Lambda -CDM$ model that our
knowledge of the present energy state of the universe would be considerably
improved with tighter constraints on the primordial dark matter $\Omega
_{dm,old}$. One unique tool to narrow down such constraints is the cosmic
microwave background (CMB) radiation. The anisotropies in CMB are a powerful
window into the early universe providing us with information about structure
formation and cosmology in general$^{25}$. CMB has this ability because the
amplitudes of its power spectrum (particularly the first peak) depend on the
amount of dark matter that was available in the early universe to create
gravitational potential wells into which the baryonic matter eventually
collected. Moreover the heights of such peaks$^{26}$ in the power spectrum
are not significantly affected by any (dark) matter $\Delta \Omega _{m}$
that might result from the effects of the recent domination of the universe
by the vacuum ($>1000$ redshifts later). The only effect that such ``new''
matter will have on the CMB power spectrum is to shift, slightly, the
horizontal (multipole) axis because of the expected $\Delta \Omega _{m}$'s
lensing effects and the resulting change in the angular distance
relationship. With the new combined data from BOOMERaNG, Maxima and DASI,
the best fit simulations$^{27}$ seem to suggest a primordial dark matter
content with a central value of $h^{2}\Omega _{dm,old}\simeq 0.1$.

Using this prior in our model (Figure 1b) we find a vacuum domination
crossing redshift VDCR of $z_{c}\simeq 1.6$. Following the same arguments as
before we see that the amount of matter produced from the decaying vacuum
since $z_{c}$ would be $\Delta \Omega _{m}\simeq 0.18$. Out of this as much
as $\left( \tilde{\Omega}_{dm,0}-\Omega _{dm,old}\right) \simeq 0.08$ would
have already deposited on clusters, leaving as much as $\left( \Omega
_{dm,0}-\tilde{\Omega}_{dm,0}\right) \simeq 0.1$ unbound in the
inter-cluster voids. This gives a vacuum dominated universe with $\Omega
_{\Lambda ,0}\simeq 0.59$ and $\Omega _{m,0}\simeq 0.41$. Figure 1b (dotted
vertical line) shows that these results, using as a prior $\Omega _{dm,old}$
associated with CMB best fit, fall well within the bounds described by the
model. On the basis that CMB is our best probe of the early universe so far,
we take these CMB-based results as our central values.

\subsection{Features and tests of the model}

In closing this discussion it is worthwhile to ask what new features the $%
D\Lambda -CDM$ model introduces into cosmology and astrophysics. We start by
noting that the model we have presented independently depicts a universe
which is presently vacuum dominated. In Section 3 we discussed the
asymptotic behavior of the equation of state for the dynamical and
interacting vacuum fluid used in our model. Such behavior uniquely
introduces\ some features in the dynamics of the universe. In our discussion
we found that such an interacting vacuum should be defined by an equation of
state that depends not only on the vacuum energy density but also generally
on the interactions between the vacuum and matter fields. In the limit of
vanishing matter fields (or weak field limit) we showed that the usual,
locally Lorentz invariant, vacuum equation of state $p=-\rho $ is recovered.
This result is usefull in as far as it suggests that the model meets this
requirement of general relativity. Moreover, as the vacuum becomes dominant
and accelerates the universe in the process, the matter production rate from
the decaying vacuum also increases. Provided the matter creation window $%
\varepsilon $ is large enough, this in turn increases the amount of matter
in the universe, effectively decreasing the vacuum to matter density ratio.
In the process, the equation of state adjusts to produce less and less
matter as the matter component of the universe starts to dominate. Such a
newly matter dominated universe would then begin to decelerate. At the same
time the matter production from the decaying vacuum becomes suppressed due
to matter dominance in the equation of state, the matter fields are
redshifting with the scale factor as $a\left( t\right) ^{-3}$ which is again
faster than the dynamical (false) vacuum that is redshiting as $a\left(
t\right) ^{-2}$. Thus soon or later the matter field density $\rho _{m}$
falls below the vacuum energy density $\rho _{\Lambda }$. At this point the
universe re-enters a vacuum dominated phase, begins to accelerate again and
to produce more matter in the process. This a state of affairs then depicts
a universe that may be self-tuning, periodically entering and exiting
inflationary phases. Such a self-tuning universe would be stable to
vacuum-induced runaway accelerations.

Our model of a universe with interacting fluids also does suggest a
conceptual explanation to the Coincidence Problem. The Coincidence Problem
expresses our lack of a simple explanation to why the energy densities in
the vacuum and matter are presently comparable$^{1}$, especially in view of
the fact that recombination $\left( z\sim 1000\right) $ and structure
formation could only have been possible if the early universe was
excessively matter dominated. In $D\Lambda -CDM$\ the matter fields redshift
faster than the dynamical vacuum by $\frac{\rho _{m}}{\rho _{\Lambda }}\sim
a\left( t\right) ^{-1}$. If matter was in dominance at some point in the
past-then, independent of the initial conditions, the two field densities
eventually cross at $z_{c}$\ as the vacuum takes dominance. We have seen in
the previous section that the combined effects of matter production from the
vacuum and the perturbations in redshifting of the fields,\ forces the
latter to interact within a decaying envelope. As the universe expands, the
amplitudes of the vacuum and the matter field densities then continue to
approach each other as they also decrease. This approach may or may not have
a local oscillatory character in time. If the fields have an oscillatory
character then there will be repeated local periods in time when their
amplitudes are comparable. In general, the above combined effects eventually
make the fields comparable as the universe evolves towards some attractor
stable state in the future. The answer to the ``why now...?'' part of the
Coincidence Problem may then be ``....because the universe is old enough.''

There are some features of the $D\Lambda -CDM$ which are potentially
verifiable. The first has to do with the suggestion the model makes that the
total matter density of the universe is more than that bound in clusters,
with the excess, $\left( \Omega _{dm,0}-\tilde{\Omega}_{dm,0}\right) $,
still unbound voids. Such smoothly distributed inter-cluster matter may be
distinguished from the vacuum by use of candles like Type Ia supernovae$^{1}$%
. However, noting (from the priors used) that the predicted amount of smooth
inter-cluster matter is only as much as $0.10\leq \left( \Omega _{dm,0}-%
\tilde{\Omega}_{dm,0}\right) \leq 0.13$ its presence could still be hidden
within the error bars (see for example (Perlmutter 1999)) of current
detection techniques. It is interesting to ask what our results look like if
we hide the above smooth dark matter within the vacuum. Recall that in our
model (and based on the priors used) we found a vacuum density $0.57\leq
\Omega _{\Lambda ,0}\leq 0.60$\ and a total (including $\Omega _{B}$) matter
density $0.40\leq \Omega _{m,0}\leq 0.43$. Clearly hiding $\left( \Omega
_{dm,0}-\tilde{\Omega}_{dm,0}\right) $ elevates the vacuum density to $%
\Omega _{\Lambda ,0}\simeq 0.7$\ while equally suppressing the matter
density to $\Omega _{m,0}\simeq 0.3$. Coincidentally, these values
correspond to those of some currently popular cosmological models.

There is, however, a second feature of $D\Lambda -CDM$ related to
inter-cluster matter that may be more readily testable. This has to do with
the suggestion we make that the newly produced dark matter in $\Delta \Omega
_{m}$ should readily respond to the long range gravitational potentials of
clusters by moving directly towards the same clusters. As a consequence,
observations of dark matter halos of clusters may fail to detect a simple
clear cut-off in the radial component of the halo density profiles.\ Recent
observations$^{28}$\textbf{\ }of matter distribution in galaxies based on
lensing \ show that density profiles of the dark matter halos manifest no
obvious cut-off, up to $300h^{-1}\,kpc$. While such preliminary results may
not constitute a vindication of the $D\Lambda -CDM$ model they tend to favor
rather than disfavor the model. A concrete verification of these predictions
can only be achieved through extensive exploration of halos. It is hoped
that future long range lensing projects$^{28}$\textit{\ }will shade some
more light on this issue\textit{.}

In the event that a fairly significant amount of \ $\Delta \Omega _{m}$ may
have accreted on to cluster dark matter halos then one more effect could be
verified in the future, with more precise measurement techniques. In that
case primordial (pre-recombination) dark matter $\Omega _{dm,old}$ will have
a deficit with respect not only to the total present dark matter content $%
\Omega _{dm,0}$ but also to that observed$^{24}$ bound-in-clusters dark
matter $\tilde{\Omega}_{dm,0}$. Such information would already be imprinted
on CMB which is\textit{\ }powerful probe into the early universe. One
recalls that in adiabatic inflationary models$^{17}$ primordial dark matter
is essential for creating the gravitational potentials that the baryons
eventually get drawn into, after recombination. This is the process of
structure formation. The freely streaming photons, after last scattering,
then carry all this information forever as amplitude perturbations in their
power spectrum. In particular, the height of the first peak in the CMB power
spectrum is very sensitive to the available amount of primordial dark matter 
$\Omega _{dm,old}$. On the other hand CMB does cleanly discriminate between
pre- and post-recombination events\textit{.} In particular, the height of
the first peak would not be affected by any dark matter $\Delta \Omega _{m}$
created long after recombination and structure formation. The $\Delta \Omega
_{m}$'s only possible effect on the CMB\ power spectrum would be to slightly
shift the curve horizontally (along the multipole moments axis) because of
the slight changes in the angular distance relation resulting from the
lensing effects. Thus the effect to look for is that CMB\ data should
consistently demand best fits that use slightly less dark matter $\Omega
_{dm,old}$ than that traditionally observed $\tilde{\Omega}_{dm,0}$ in
galaxies and clusters, by such standard techniques as X-ray surveys. It has
been pointed out before by several experts (see for example$^{29}$) that CMB
fits may be improved by initial (pre-recombination) low dark matter values.
The recent CMB data, for example, seems to suggest best fits$^{29}$ with
central values of $\Omega _{dm}h^{2}=0.1^{+.07}$. However even with central
values as low as these the results still lie within the error bars of the
usually quoted dark matter values. Thus, if the effect exists, it can only
be isolated by future higher precision measurements.

\section{Conclusion}

In the preceding discussion we have investigated the possible effects of a
decaying vacuum energy density provided by a dynamical positive cosmological
constant $\Lambda \left( t\right) $. On applying generalized energy
conservation conditions one finds that such vacuum decay will produce matter
fields. This matter production by a dynamical cosmological constant is
triggered by the recent domination of the universe by vacuum energy. Our
model also assumes inflation and big bang nucleothenthesis (BBN). With these
assumptions we have written down several of the cosmological quantities,
including the present fractional densities of matter $\Omega _{m,0}$ and
vacuum $\Omega _{\Lambda ,0}$ contents in the universe in terms of only one
parameter, namely: the recent vacuum domination crossing redshift (VDCR), $%
z_{c}$. This dependence on $z_{c}$ makes these densities observable
quantities. A central feature of this model is that the amount of matter $%
\Omega _{m,0}$ in the current universe is more than that $\Omega _{m,old}$
at the time of recombination by an amount $\Delta \Omega _{m}$ that has been
created since vacuum domination at $z_{c}$. While this ``new'' matter is
produced smoothly everywhere it could accrete onto clusters through the
long-range gravitational potentials of the latter.

To relate our arguments to observation we first estimate when the vacuum
comes to dominate the universe, by constraining $z_{c}$. This is done by
establishing reasonable bounds on the primordial dark matter, $\Omega
_{dm,old}$. In our calculations we apply only the central values of such
bounds. Given these priors we constrain the vacuum domination crossing
redshift to $0.89\leq z_{c}\leq 2.0$. Within these constraints we find that
from $z=z_{c}$\ to the present, the decay of dark energy should have created
matter to the tune of \ $0.13\leq \Delta \Omega _{m}\leq 0.20$ with as much
as $0\leq \left( \tilde{\Omega}_{dm,0}-\Omega _{dm,old}\right) \leq 0.10$\
already deposited onto the clusters, leaving as much as $0.10\leq \left(
\Omega _{dm,0}-\tilde{\Omega}_{dm,0}\right) \leq 0.13$\ to be inter-cluster
unbound matter. Consequently, within these constraints, the $D\Lambda -CDM$\
model finds that the universe is presently vacuum dominated with a vacuum
density of \ $0.57\leq \Omega _{\Lambda ,0}\leq 0.60$\ and a total matter
density of $0.40\leq \Omega _{m,0}\leq 0.43$. Further, using as a prior the
dark matter density that currently gives the best fit to the latest CMB
data, we find that the results (Figure1, dotted vertical line) fall well
within the boundaries we already established. When used as a prior the CMB
based $h^{2}\Omega _{dm,old}\simeq 0.1$ predicts in our model a universe
that has been vacuum dominated since $z_{c}\simeq 1.6$ and currently has a
total matter density of $\Omega _{m,0}\simeq 0.41$ and a vacuum dark energy
density of $\Omega _{\Lambda ,0}\simeq 0.59$. We take these results as our
central values.

In closing we summarize some of the new features the model introduces in
cosmology. First, the model points to an existence of newly formed dark
matter in the inter-cluster voids and suggests that some of this matter
should already be accreting onto clusters through the latter's long range
gravitational potentials. There are two consequences of this. If some of the
newly produced matter is indeed already leaking onto the cluster then the
latter's dark matter halos will be extended in space and demonstrate little
or no clear cut-off in their radial density profiles. Secondly, if indeed a
substantial amount of this matter has already landed then the amount of dark
matter we observe bound in clusters will be greater than that needed to
produce the best CMB fits by an amount that might have already deposited
onto the clusters. The difference between these two values, if it exists,
could still be hidden within the measurements' error bars. A final feature
of our model relates to the combined effect of the matter production by a
decaying vacuum and the different rates namely, $a\left( t\right) ^{-3}$ and 
$a\left( t\right) ^{-2}$ at which matter and the vacuum respectively dilute
with the scale factor. Such combination may create conditions for a universe
in which the vacuum and matter densities evolve towards comparable and
decreasing amplitudes. This suggests a conceptual explanation to the
Coincidence Problem. It also implies that the universe \textit{may} be
stable to vacuum-induced runaway accelerations.

In order to establish the validity of such notions, it is evident that more
work needs to be done including developing exact functional form of the
equation of state. Such efforts are being undertaken and the results will be
reported elsewhere.

\begin{acknowledgement}
This work was made possible by funds from the University of Michigan
\end{acknowledgement}

\end{document}